\documentclass[aip]{revtex4-1}

\usepackage{graphicx}
\usepackage{amsmath}
\usepackage{amsfonts}
\usepackage{amssymb}
\usepackage{float}
\usepackage{empheq}
\usepackage{graphicx}
\usepackage{subfig}
\usepackage{gensymb}
\usepackage{amsthm}
\usepackage{siunitx}
\usepackage{makecell}
\usepackage[verbose]{placeins}
\usepackage{blindtext}
\usepackage{scrextend}
\usepackage{tablefootnote}
\usepackage{blindtext}

\begin{document}
\title{Optically controlled dual-band quantum dot infrared photodetector}

\author{Stefano Vichi}
\email{stefano.vichi@unimib.it}
\affiliation{INFN, Sezione di Milano-Bicocca, I-20126 Milano, Italy\\}
\author{Sergio Bietti}
\affiliation{L-NESS and Department of Materials Science, University of Milano-Bicocca,  via Cozzi 55, 20125 Milan, Italy\\}
\author{Francesco Basso Basset}
\affiliation{Department of Physics, Sapienza University of Rome, Piazzale A. Moro 5, I-00185 Rome, Italy} 
\author{Artur Tuktamyshev}
\affiliation{INFN, Sezione di Milano-Bicocca, I-20126 Milano, Italy\\}
\author{Alexey Fedorov}
\affiliation{L-NESS and CNR-IFN, 20133 Como, Italy\\}
\author{Stefano Sanguinetti}
\affiliation{INFN, Sezione di Milano-Bicocca, I-20126 Milano, Italy\\}
\affiliation{L-NESS and Department of Materials Science, University of Milano-Bicocca,  via Cozzi 55, 20125 Milan, Italy\\}

\maketitle

\section{Abstract}

We present the design for a novel type of dual-band photodetector in the thermal infrared spectral range, the Optically Controlled Dual-band quantum dot Infrared Photodetector (OCDIP).
This concept is based on a quantum dot ensemble with a unimodal size distribution, whose absorption spectrum can be controlled by optically-injected carriers. An external pumping laser varies the electron density in the QDs, permitting to control the available electronic transitions and thus the absorption spectrum. We grew a test sample which we studied by AFM and photoluminescence. Based on the experimental data, we simulated the infrared absorption spectrum of the sample, which showed two absorption bands at 5.85 $\mu$m and 8.98 $\mu$m depending on the excitation power.

Keywords: infrared photodetector, quantum dot, droplet epitaxy

\section{Introduction}

Thermal infrared (TIR) remote sensing plays a key role in different fields, which span military, commercial, public and academic domains and it is of particular interest in Earth science research, where it permits the modeling of surface thermal energy \cite{Hulley2010}. TIR remote sensing is based on the detection of objects infrared emission at multiple wavelengths in order to eliminate spurious effects and reconstruct the real object temperature.
These wavelengths, which are typically located between 8 and 13 $\mu$m, are chosen in the spectral region with the highest atmospheric transmissivity. A two-channel (or split-window) algorithm requires at least two absorption bands in the TIR region \cite{Dozier1981,Sobrino2014} in order to determine the absolute temperature of the studied surface. This wavelength restriction can be avoided if the distance from the radiation source is sufficiently low to make the atmospheric absorption negligible.
This opens up the possibility to detect temperatures in the 200-300 $\degree$C range, which corresponds to an emission peak of approximately 5.5 $\mu$m.
Measuring multiple wavelengths typically requires different detectors with separate cooling systems and electronics. However, the difficulties and costs of assembling several detectors can be overcome by using a single detector responding in multiple bands. For this reason, dual-band quantum well infrared photodetectors (QWIPs) have been extensively researched and significant progress has been made in the past years \cite{Jiang1999}. In recent times, quantum dots (QDs) infrared photodetectors (QDIPs) \cite{Downs2013,MarRog08,Martyniuk2009} have emerged as a promising alternative to QWIPs. Compared to these, QDIPs offer numerous advantages, including sensitivity to normally incident light, low leakage current and low dark current \cite{Sanguinetti2008}. In order to detect multiple wavelengths, most of the reported multicolor QDIPs incorporate QDs with different sizes \cite{Rogalski2002, Downs2013}. Due to the three-dimensional quantum size effect, the TIR detection bands can be tuned in the 5-12 $\mu$m range. In state-of-the-art dual-band  QDIP,  two QD ensembles with different sizes are present and thus different detection bands can be selected.  
Here we present the idea for a novel type of dual-band photodetector in the TIR spectral window, the Optically Controlled Dual-band Infrared Photodetector (OCDIP), which avoids the disadvantages of having multiple QDs ensembles and at the same time allows the dynamical tuning of the optical response for band contribution separation. The OCDIP design is based on a single QD ensemble whose absorption spectrum can be controlled by optically injected carriers. An external pumping laser varies the electron density in the QDs, permitting to control the available electronic transitions and thus the absorption spectrum. In order to be achievable, it is fundamental that the two transitions are distinguishable and that those states can be filled selectively.
In this work we also propose a geometry of the nanostructures in order to have two absorption bands at 5.85 $\mu$m and 8.98 $\mu$m. The proposed structure would be particularly suitable for the detection of forest fires, since the wood ignition temperature of approximately 250 $\degree$ C corresponds to an IR emission peak of 5.5 $\mu$m.

\section{OCDIP concept}

In order to fill the QDs ground state, usually a $\delta$-doping layer is grown in the proximity of the QDs layers \cite{Pan1998}. Electrons are thermally excited from donor states into QDs, populating them and contributing to the absorption process. However, since the doping density is fixed from the growth conditions, it is not possible to dynamically control the electron density inside QDs and thus the absorption spectrum. Furthermore, there are several problems related to the $\delta$-doping layer, such as potential fluctuations, random distribution of dopants, leakage current paths \cite{Liu2003}, internal electric field \cite{Yakimov2012} and increase in dark current \cite{Attaluri2006}. Our approach to fill QDs ground state is designed to enable the dynamical control of the QDs electron population. In the OCDIP approach, electron-hole pairs are generated in the barrier due to optical excitation and then are captured by QDs. The electron density inside QDs is proportional to the optical pumping power and therefore a change of the incident power affects QDs electron population. 

Usually more than two confined states are present in QDs; if only the QD ground state is populated, a single intraband absorption peak is present, corresponding to the difference in energy between the electron ground state and the continuum. A second absorption band, originating from the transition of electrons confined in the first excited state, can be accessed by increasing the carrier population inside the QDs enough to saturate its ground state. After that, the first excited state starts to fill, enabling also the transition from the first excited state to the continuum. The control over QDs electron population, on which is based the OCDIP idea, is thus the key to selectively obtain one or two absorption bands and dynamically switch between them. Höglund et al. demonstrated that resonant optical pumping across bandgap can be used to artificially dope QDs and selectively obtain dual-band absorption in a dots-in-a-well infrared photodetector (DWELL) \cite{Hoglund2009}. However, this system relies on resonant excitation and therefore needs two high-power lasers of different wavelengths to pump electrons in the ground and first exited states of the QDs respectively. Due to the low absorption of the resonant excitation, this is not a particularly efficient system to dynamically control the electron population in the QDs. These difficulties can be overcome by combining the idea of Höglund with the one of Ramiro \cite{Ramiro2015}, who recently showed an optically triggered single-band IR photodetector. In such a design the absorption of the TIR band is triggered by an external light source which injects carriers in the QDs. Their idea of filling the ground state by optical injection can be further extended to obtain dynamical control of QDs electron density, in order to achieve the dual-band photoresponse described in this paper.

\section{Experimental}

In order to demonstrate the feasibility of this concept, we grew GaAs QDs in an Al$_{0.3}$Ga$_{0.7}$As barrier by droplet epitaxy\cite{Watanabe2000,Gurioli2019}. We chose this technique over the more common Stranski-Krastanov since it is the one that allows the best control over the nucleation process and to achieve very low size dispersion\cite{Bietti2015,BassoBasset2019} which leads to a narrow absorption line \cite{Vichi2020}.

The sample was prepared in a conventional III-V molecular beam epitaxy machine equipped with cracked valved As cell and in-situ reflection high energy electron diffraction. The samples were grown on a GaAs (001) 2" substrate. After the deposition of an Al$_{0.3}$Ga$_{0.7}$As layer acting as a potential barrier for the QDs, Ga droplets were formed by irradiating the surface with a flux of 0.02 ML/s at a temperature of 300°C. The first monolayer of Ga reacts with the As-rich c(4x4) reconstructed surface establishing a Ga-stabilized (4x6) reconstruction. The droplets are then formed by the remaining Ga coverage, resulting in the deposition of 0.06 MLs. Then the Ga droplets were exposed to an As beam equivalent pressure of 5·10$^{-5}$ torr at 150°C for 3 minutes to crystallize into GaAs. Subsequently, the nanocrystals underwent a flash procedure, consisting of a 10 minutes heating at 380°C in an As pressure of 4·10$^{-6}$ torr. Finally, the QDs were covered with another layer of 10 nm of Al$_{0.3}$Ga$_{0.7}$As deposited at low temperature followed by 140 nm at 580°C, and capped with 10 nm of GaAs. After the growth, the sample underwent a rapid thermal annealing in nitrogen atmosphere at 750°C for 4 minutes. 

Photoluminescence measurements (PL) were performed at different temperatures and powers in order to study the filling of QDs. 
During photoluminescence (PL) measurements the sample was kept in a closed-cycle cryostat with the temperature set stable by a heater and a PID controller. The QDs were excited using the emission at 532 nm of a Nd:YAG laser focused on a spot with a diameter of about 80 $\mu$m. The PL signal was collected by a spherical mirror and sent into a 500 mm focal length spectrometer equipped with a 150 gr/mm grating and a Peltier-cooled CCD. 
The experimental results were then compared with numerical simulations based on an eight bands \textbf{k}$\cdot$\textbf{p} model performed using the commercial software TiberCAD \cite{TiberCAD}. Finally, the intraband spectrum was simulated based on the analysis of the PL results.

\section{Results and Discussion}

\begin{figure}
	\centering
	\includegraphics[width=\linewidth]{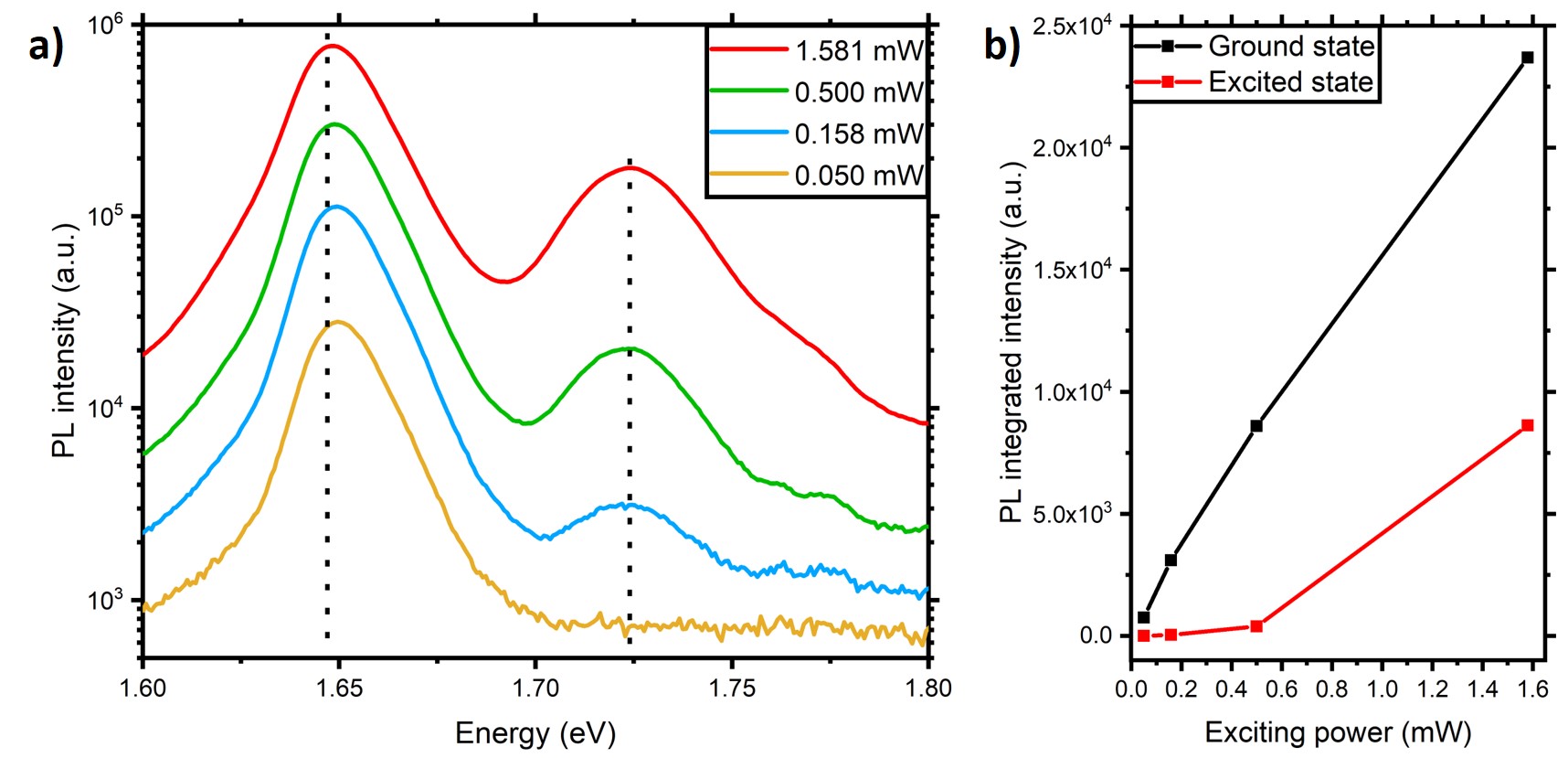}
	\caption{a) Photoluminescence spectrum as a function of incident power measured at 90 K. The dashed lines indicate the simulated transition energies of the ground state transition (1.647 eV) and of the first excited state transition (1.724 eV). b) Photoluminescence integrated intensity of the ground and first excited states transitions as a function of the incident power.}
	\label{fig:pl}
\end{figure}

In figure \ref{fig:pl}a) are shown the PL spectra taken at 90 K of the selected sample for different values of the incident power. As can be seen, at the lowest power only the ground state transition of the QDs ensemble is visible. By increasing the excitation power, also the excited states become populated and therefore new transitions appear. The black dotted lines indicate the simulated transition energies of the ground state and of the first excited states. 
In the panel b of the same figure are reported the PL integrated intensities of the ground and the first excited state as a function of the excitation power. As can be seen, the intensity of the ground state emission increases linearly with increasing excitation power. On the other hand, the emission intensity of the first excited state increases more than linearly.
This demonstrates that it is possible to control the population of each confined state in a precise way by changing the optical excitation power. 
In this work we are interested in the transitions involving the first two electron states at lower energies (i.e. ground state and first excited state). From the figure it is possible to see that 0.050 mW (corresponding approximately to 1 $W cm^{-2}$) is the highest power for which only the ground state is filled. On the other hand, 0.500 mW (corresponding approximately to 10 $W cm^{-2}$) is the largest power for which also the second state is populated, while the third state is still empty. 
Since the emission intensity is proportional to the filling of the states, by comparing the integrated intensities of the transitions it is possible to estimate the relative population density of the states at the selected powers and use it as an input parameter to simulate the intraband absorption spectrum.

We performed numerical simulations using an eight bands \textbf{k}$\cdot$\textbf{p} model and envelope function approximation in order to compute the transition energies and the optical matrix elements of those transitions. The simulated structure consists of a truncated cone QD made of GaAs with major and minor radius of 8 nm and 3.6 nm respectively \cite{Bietti2015} and a height of 8.2 nm, as obtained from the analysis of the AFM images, included in an Al$_{0.3}$Ga$_{0.7}$As barrier.
The band-to-band transition between electrons and heavy holes states are indicated in figure \ref{fig:pl}a) by dashed lines and show a good agreement with the experimental results. The calculated transition energies are 1.647 eV and 1.724 eV for the ground and the first excited state respectively.
The computed optical matrix elements $|M|^2$ for these transitions are 0.6414 a.u. and 0.0327 a.u. for the ground state and the first excited state transitions.
The integrated intensity of a PL peak can be expressed as the product of a power-dependent effective carrier density and the optical matrix element:

\begin{equation}\label{eq:carrier}
	I^{PL}_{GS(ES)}(P) = N^{eff}_{GS(ES)}(P)  |M_{GS(ES)}|^2
\end{equation}

in which $I^{PL}_{GS(ES)}(P)$ is the PL integrated intensity shown in figure \ref{fig:pl}b) of the ground (excited) state peak, $ N^{eff}_{GS(ES)}(P)$ is the power-dependent effective carrier density of the ground (excited) state and $|M_{GS(ES)}|^2$ are the optical matrix elements of the transitions.
Therefore, by deconvoluting the PL spectra of figure \ref{fig:pl}a) into its ground and excited state components, computing their integrated intensities (figure \ref{fig:pl}b)) and applying equation \ref{eq:carrier}, it is possible to obtain the values of $N^{eff}_{GS(ES)}(P)$ at a chosen power.
These values can than be used in the analysis of the IR photocurrent signal, in order to separate the contribution of the ground state and the first excited state transitions.

\begin{figure}
	\centering
	\includegraphics[width=\linewidth]{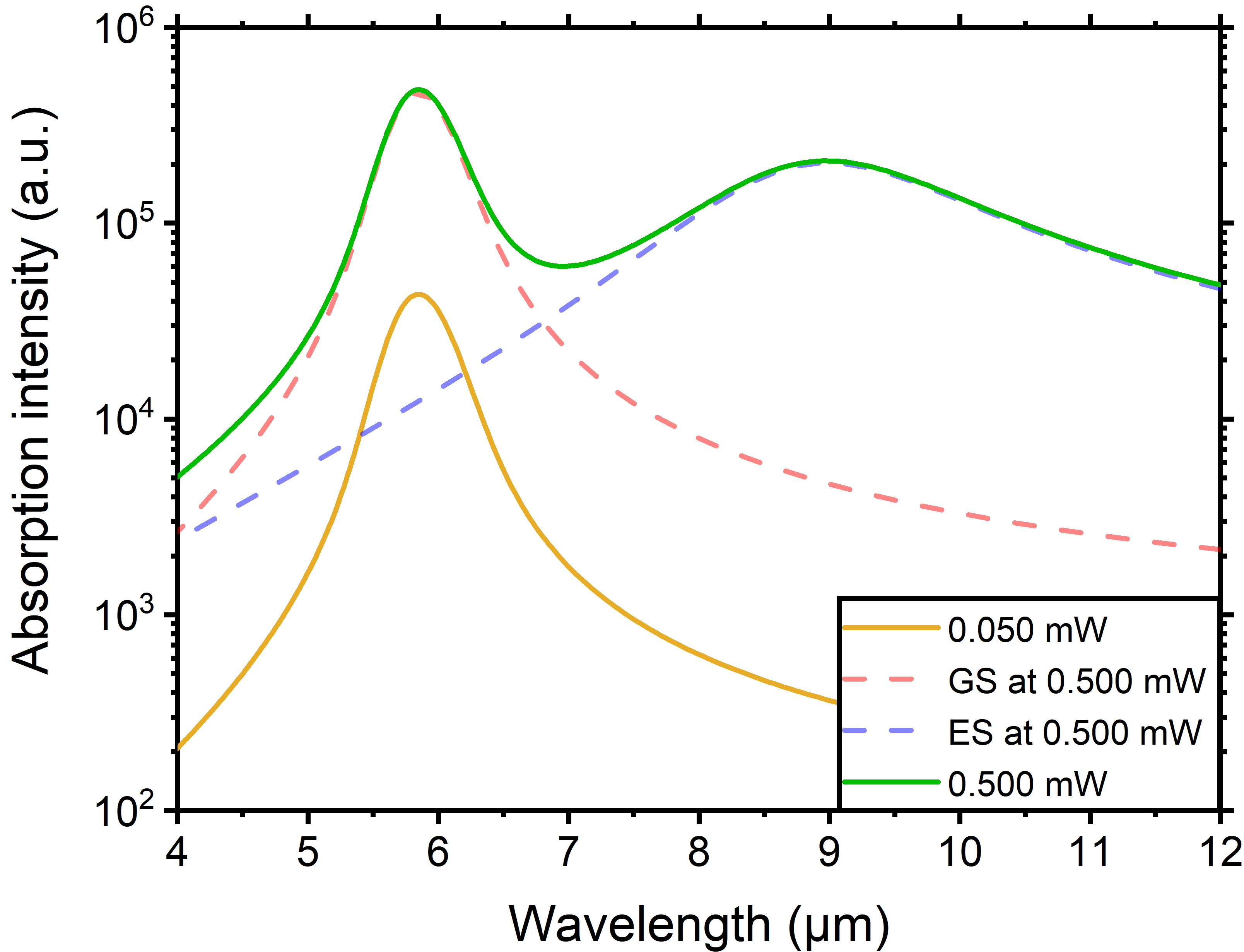}
	\caption{Simulated intraband absorption spectrum for ground and first excited state at a pumping power of 0.050 mW and 0.500 mW and a temperature of 90 K. The dashed lines show the simulated absorption spectra of the ground and the first excited state at a pumping power of 0.500 mW.}
	\label{fig:abs}
\end{figure}

Based on the simulations, the obtained ground state and first excited state electron energies are at 0.212 eV and 0.138 eV below the Al$_{0.3}$Ga$_{0.7}$As barrier energy, corresponding to the transition wavelengths of 5.85 $\mu$m and 8.98 $\mu$m respectively.
To simulate the absorption spectrum from the transition energies, we assumed the intraband absorption peak to have the same shape as the PL emission peaks (i.e. a Voigt function). For these transitions we considered the FWHM to be half of the one measured in the PL spectra (figure \ref{fig:pl}a)). This choice is based on the assumption that both electrons and holes states contribute to the broadening of the PL spectrum in a similar way. As a consequence, in first approximation we expect the FWHM to be halved for an intraband transition, where only the conduction band is involved.
The intensity of the intraband absorption was then calculated using the effective carrier density obtained from equation \ref{eq:carrier} and assuming that the IR absorption is proportional to the filling of those states. Since the final state of the transitions that we are considering is a delocalized conduction band state (i.e. Bloch wave), we assume the optical matrix elements to be similar for both ground and first excited state.
The resulting absorption spectrum is shown in figure \ref{fig:abs}. In particular, in the case of 0.500 mW, the red and blue dashed lines show the contributions of the ground and the first excited state respectively.

As can be seen, for an exciting power of 0.050 mW a single absorption band peaking at 5.85 $\mu$m is expected, while for 0.500 mW a second band at 8.98 $\mu$m is present due to the filling of the first excited state. In the latter case, it is possible to note that the overlap between the two transitions is quite low.
However, in an actual device being able to separate the contribution of the two transitions when measuring the photocurrent requires some initial calibrations. In the following, we will discuss a procedure to be able to determine the signal of each individual band.

\section{Detection Mechanism}

\begin{figure}
	\centering
	\includegraphics[width=0.5\linewidth]{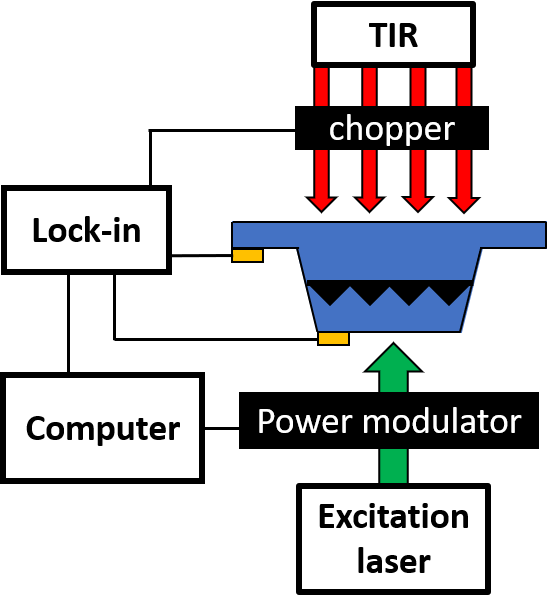}
	\caption{Schematics of the proposed device and its detection mechanism.}
	\label{fig:device}
\end{figure}

In order to have a working detector that combines two absorption bands in the same structure, it must be possible to distinguish the origin of each signal.
One possibility is to use a focal plane array with a grating that separates the spectral components of the incident light. However, this method can not be used for a single pixel device. In this second case, the identification of the signal can be done by alternating the optical pumping power between 0.500 mW and 0.050 mW while continuously measuring the photocurrent. After an initial calibration, an unknown signal can then be decomposed into its components.
This idea is based on the fact that the electron population inside the QDs is determined only by the pumping power, assuming the steady state behavior. This assumption is reasonable since, after a small transient, the system will be at the equilibrium as happens during PL measurements. Therefore, the signal can be deconvoluted into its components using a procedure analogous to equation \ref{eq:carrier}. 
Analyzing the PL spectrum gives information about the relative population of the states at the selected powers. Once this value is obtained, the photocurrent spectrum has to be measured by alternating the pumping powers in order to constantly change the QDs population density between the chosen values. At 0.050 mW the photocurrent is determined only by the absorption band peaking at 5.85 $\mu$m. When the excitation power is changed to 0.500 mW, the photocurrent originated from the ground state and the excited state transitions can be calculated by considering the new population densities of those states.
Figure \ref{fig:device} shows the schematics of the proposed device. The excitation laser reaches the sample from the top side of the mesa, so that the photo-generated carriers can reach and fill the QDs layer. A power modulator controlled by a computer continuously changes the incident excitation power between 0.050 mW and 0.500 mW in order to control the filling of the QDs states.
The IR radiation incident from the bottom of the device is chopped and the photocurrent signal is then read by a lock-in amplifier and sent to the computer for the analysis.
Thanks to the chopper and the power modulator, it is possible to distinguish the IR signal from the dark current in real-time at both excitation powers. 
Moreover, by performing a similar analysis to the one described above it is possible to separate the contribution of the two absorption bands.

\section{Conclusions}

In this work we have proposed a new idea for a dual-band infrared photodetector, the Optically Controlled Dual-band Infrared Photodetector (OCDIP). By optically controlling the electron density inside the QDs, it is possible to obtain selectively one or two absorption bands. We demonstrated the concept by growing a sample and measuring the PL spectrum at different operation powers of a visible laser. Based on the grown sample and the results of PL measurements, we simulated the intraband absorption spectrum at two selected power settings, namely 0.050 mW and 0.500 mW. The result shows the possibility to have one or two separated absorption bands which can be controlled by changing pumping power. This theoretical simulation of an optically controlled dual band photodetector may open new path for low cost and compact devices.

\section{Acknowledgement}
We acknowledge funding by PIGNOLETTO project, co-financed with the resources POR FESR 2014-2020, European regional development fund with the contribution of resources from the European Union, Italy and the Lombardy Region.

Francesco Basso Basset acknowledge financial support from the European Research Council (ERC) under the European Union’s Horizon 2020 Research and Innovation Programme (SPQRel, Grant Agreement No. 679183) and by the European Union’s Horizon 2020 Research and Innovation Program under Grant Agreement  No. 899814 (Qurope).

\section{Author Contributions}
Conceptualization, Stefano Vichi; Data curation, Sergio Bietti, Francesco Basso Basset, Artur Tuktamyshev and Alexey Fedorov; Formal analysis, Stefano Vichi and Francesco Basso Basset; Funding acquisition, Stefano Sanguinetti; Investigation, Sergio Bietti, Francesco Basso Basset, Artur Tuktamyshev and Alexey Fedorov; Methodology, Stefano Vichi; Supervision, Stefano Sanguinetti; Validation, Stefano Sanguinetti; Writing – original draft, Stefano Vichi.

\bibliography{OCDP}
\end{document}